\documentstyle[emulateapj]{article}

\slugcomment{To appear in Astronomical Journal}

\lefthead{Plana et al.}

\righthead{Kinematics and morphology of the ionized gas in 
the compact group of galaxies HCG 18}

\input epsf.sty
\begin{document}

\title{Kinematics and morphology of ionized gas in Hickson Compact Group 18}

\author{H. Plana\altaffilmark{1,2}}
\affil{Observatorio Astronomico Nacional, UNAM, Apartado Postal 877, 22800,
Ensenada, BC, Mexico}

\author{P. Amram \altaffilmark{2}}
\affil{IGRAP, Observatoire de Marseille,
       2 Place Le Verrier, 13248 Marseille Cedex 04, France }

\author{C. Mendes de Oliveira\altaffilmark{3}}
\affil{Instituto Astron\^omico e Geof\'{\i}sico (IAG),
       Av Miguel Stefano 4200 CEP: 04301-904 S\~ao Paulo, Brazil}

\and

\author{C. Balkowski \altaffilmark{4}}
\affil{Observatoire de Paris, DAEC, UMR 8631, CNRS et Universite Paris 7, 
       F-92195 Meudon Cedex, France}



\begin{abstract}

We present new observations of $H\alpha $ emission in the Hickson Compact
Group 18 (HCG 18) obtained with a scanning Fabry-Perot interferometer. The
velocity field does not show motions of individual group members but,
instead, a complex common velocity field for the whole group.  The gas distribution is very asymmetric with clumps of
maximum intensity coinciding with the optically brightest knots. 
Comparing $H\alpha $ and HI data we conclude that HCG 18
is not a compact group but instead a large irregular galaxy with several clumps of star formation. 

\end{abstract}


\section{Introduction}

Compact groups of galaxies have been known for over 30 years (Vorontsov-Vel`yaminov 1959, Arp 1966, Rose
1977, Hickson et al. 1977, Hickson 1982).  A spectroscopic survey
confirmed that 92 of the 100 groups catalogued by Hickson (1982) have at
least three accordant redshift members with 69 groups having at least four (Hickson et al. 1992).  With a median galaxy-galaxy separation
of 40 h$^{-1}$ kpc  and  a low typical velocity dispersion ($\sigma$
$\sim$ 200 km s$^{-1}$, Hickson et al. 1992), compact groups are usually
considered to be ideal laboratories for studying galaxy interactions.
The most direct way to determine if interactions have occurred among
group galaxies is to measure their kinematics in order to check if
they are disturbed or normal. Study of the ionized gas kinematics
for several of the Hickson group galaxies has shown that they 
are in different evolutionary stages (Mendes de Oliveira et al.
1998, Plana et al.  1998).

HCG 18 (H18, Arp 258 or VV 143) was cataloged by Hickson (1982) as a
group of three irregular galaxies (H18b, c and d with radial
velocities between 4080 and 4163 km $s^{-1}$) plus one discordant-redshift
S0 galaxy (H18a with a velocity of 10000 km $s^{-1}$; Hickson 1993). In
this paper the H18 group refers to the triplet H18b, c and d.

Verdes-Montenegro et al. (1998) determined the IRAS flux for H18d and
upper limits on the flux for H18b and H18c.  They also determined the
H$_2$ mass for H18b and H18c, from their CO observations. Allam et al. (1996)
also give IRAS flux but for H18b, c and d together. We eventually use Allam's values because H18 IRAS emission can not be spatially resolved. HI data is also
available for this group, besides the optical measures (Hickson 1993).
From a study of the HI velocity field of the group Williams
and van Gorkom (1988, hereafter W\&vG) concluded that the HI gas is
concentrated in a single cloud with m$_{HI}$ = 10$^{10}$ M$_{\odot}$ encompassing all of the optical structures.

%
%

\begin{deluxetable}{lll}
\tablenum{1}
\tablewidth{0pt}
\tablecaption{Journal of Perot-Fabry observations}
\tablehead{
\colhead{} & \colhead{Compact Group of Galaxies Hickson 18 } & \colhead{}}
\startdata
Observations & Telescope & CFHT 3.6m  \nl 
	     & Equipment & MOS/FP @ Cassegrain focus  \nl 
	     & Date & August, 22th 1996  \nl 
	     & Seeing & $<$ 1" \nl 
Interference Filter &   Central Wavelength & 6653 \AA \tablenotemark{1} \nl 
		    &   FWHM & 23 \AA  \tablenotemark{2} \nl 
		    &   Transmission (maximum) & 0.6  \nl 
Calibration & Neon Comparison light & $\lambda$ 6598.95 \AA \nl 
Perot--Fabry & Interference Order & 1162 @ 6562.78 \AA \nl 
		 & Free Spectral Range at H$\alpha$ & 265 km s$^{-1}$ \nl 
		 & Finesse at H$\alpha$ & 12 \nl 
                 & Spectral resolution at H$\alpha$ & 27344 at the sample step \nl
Sampling & Number of Scanning Steps & 24 \nl 
	 & Sampling Step & 0.24 \AA\ (11 km s$^{-1}$)\nl 
	 & Total Field & 430''$\times $430'' (500$\times $500 px$^2$) \nl 
	 & Pixel Size & 0.86'' \nl 
Detector & & STIS 2 CCD  \nl 
Exposures times & Total exposure & 2 hours \nl 
		& Total exposure time per channel & 300s \nl

\enddata
\tablenotetext{} { $^1$ For a temperature de 5$^o$. $^2$ For a mean beam 
inclination of 2.7$^o$. }

\end{deluxetable}

  We observed the H18 group  in the H$\alpha$ emission line with a scanning Fabry-Perot instrument and we obtained velocity
and H$\alpha$ integrated flux maps for the system. This observation is part of a
larger program that has the goal of unveiling kinematic evidences of
interactions in compact groups of galaxies in order to determine their
evolutionary stages.

  H18 appears to be one of the most complex groups of the Hickson
sample. A key question raised by the HI study of W\&vG 
 was the nature of the three components b, c and d, namely if they are
individual entities or form one single galaxy. Based on our
observations and comparisons between our data and the HI results,
we conclude that H18 is likely to be a single large irregular galaxy. The H$\alpha$
maps presented in this study permit the determination of the H$\alpha$
luminosity (L(H$\alpha$)) and star formation rate (SFR) of the
system, which can be compared with typical values for samples
of irregular galaxies studied by Hunter et al. (1986, 1989 and 1993).
We also compare the total B luminosity (L$_B$) of the group and its
far-infrared luminosity (L($_{FIR}$)) with values given in the literature
for typical irregular galaxies.

The paper is organized as follows. Section 2 gives details about the
reduction of the Fabry-Perot data. Section 3 describes 
the results. Section 4 has a comparison of our results with the HI
observations. Section 5 presents  a discussion about the nature of H18
in light of our new H$\alpha$ observations and section 6 contains our
summary and final remarks.

\section{Observations and data reduction}

 Observations were carried out in August 1996 with the multi-object
spectrograph focal reducer in Fabry-Perot mode attached to the F/8
Cassegrain focus of the 3.6m Canada-France-Hawaii telescope (CFHT).
The CCD was a STIS 2 detector, 2048 $\times$ 2048 pixels with a
read-out noise of 9.3 e$^{-}$ and a pixel size on the sky of 0.86"
after 2x2 binning.  Table 1 contains the journal of observations.
Reduction of the data cubes were performed using the CIGALE/ADHOC
software (Boulesteix 1993).  The data reduction procedure has been
extensively described in Amram et al. (1996).

  Wavelength calibration was obtained by scanning the narrow Ne 6599
\AA\ line under the same conditions as the observations. Velocities measured relative to the systemic velocity are very accurate,
with an error of a fraction of a channel width (${\rm <3 \, km
s^{-1}}$) over the whole field.

   Subtraction of bias, flat fielding of the data and
cosmic-ray removal have been performed for each image of the data cube.
To minimize seeing variation, each scan image was smoothed with a
gaussian function of full-width at half maximum equal to the
worse-seeing data of the data cube.  Transparency and sky foreground
fluctuations have also been corrected using field star fluxes and
galaxy-free windows.

    The signal measured along the scanning sequence was separated
into two parts: (1) an almost constant level produced by the
continuum light in a 23 \AA ~passband around H$\alpha $ (continuum
map), and (2) a varying part produced by the H$\alpha $ line
(H$\alpha$ integrated flux map).  The continuum level was taken to be the mean of
the three faintest channels, to avoid channel noise effects.  The
H$\alpha$ integrated flux map was obtained by integrating the monochromatic profile in each pixel. The velocity sampling was 11 $km$ $s^{-1}$.  The
H$\alpha$ integrated flux map had one-pixel resolution in the positions of H18b, H18c and H18d.
Spectral profiles were binned in the outer parts  to 5x5
pixels in order to increase the signal-to-noise ratio.  Strong OH
night sky lines passing through the filters were subtracted by
determining the level of emission away from the galaxies (Laval et al.,
1987).

A rough flux
calibration was attempted using the group HCG 100 observed during 
two runs (1995 at the ESO 3.6m and 1996 at the CFHT). 
Monochromatic images of
HCG 100 have been calibrated in flux using the Cartwheel Galaxy,
observed during the 1995 ESO run (see Amram
et al. 1998 for details). We then determined the HCG 18 galaxy fluxes relative to HCG 100.  We have reasonable agreement with fluxes published by
Iglesias-Par\'amo \& V\'{\i}lchez 1999 for H100a and H100b.  We
compared our H18 fluxes with those determined by Laurikainen and Moles
(1989) and we found a value that was four times higher for the flux of H18d and around
five times for H18b (they did not measure the flux of
the H18c). Difference by factors of four or five can be 
understood if we note that flux calibration
with slit observations can significantly underestimate the true flux. H$\alpha$ profiles
for the H18 galaxies were measured to a minimum flux density of 1.1
$\times$ 10$^{-16}$ erg s$^{-1}$cm$^{-2}$ arcsec$^{-2}$ and a maximum
of 9.3 $\times$ 10$^{-15}$ erg s$^{-1}$cm$^{-2}$ arcsec$^{-2}$
(corresponding to a S/N between 3 and 500).

%
%

\begin{deluxetable}{llll}
\tablenum{2}
\tablewidth{0pt}
\tablecaption{Physical Parameters}
\tablehead{
\colhead{} & \colhead{HCG 18b}& \colhead{HCG 18c}& \colhead{HCG 18d}}
\startdata
Other names  & UGC 2140a & UGC 2140b & UGC 2140c \nl
	  &  Arp 258 / VV 143 & Arp 258 / VV 143 & Arp 258 / VV 143 \nl
$\alpha$ (1950) $^1$ & 02$^{h}$36$^{m}$18.5$^{s}$ &   
02$^{h}$36$^{m}$18.2$^{s}$&  02$^{h}$36$^{m}$17.0$^{s}$ \nl

$\delta$ (1950) $^1$ & 18$^o$10'04.1" & 18$^o$10'24.5" & 
18$^o$10'43.9" \nl

Morphological type (Hickson) $^1$ & Im & Im & Im \nl

B$_{T_C}$ $^1$ & 14.90  & 15.61 & 15.10 \nl

Systemic heliocentric velocity/$^1$ (km s$^{-1}$)  & 4082 $\pm$ 40
& 4163  $\pm$ 37 &  4067 $\pm$ 58 \nl

Gas central velocity (km s$^{-1}$) & 4063 $\pm$ 20  & 4108 $\pm$ 20 & 4061 $\pm$ 20   \nl

D(Mpc)$^2$ & 54 & 54  & 54 \nl

HI velocities/$^2$ (km s$^{-1}$) & 4065 & 4085 & 4095 \nl

FWHM of central profiles (km s$^{-1}$) & 65 $\pm$ 20& 70 $\pm$ 20& 80 $\pm$ 20 \nl

\enddata
\tablenotetext{} { $^1$ Hickson 1993  $^2$ Williams \& van Gorkom 1988}

\end{deluxetable}

\section{Results}

We obtained H$\alpha$ integrated flux map and velocity maps for H18b, c and d.
The systemic velocity of H18a was outside of the range of the interference
filter we used. Fig. 1 shows continuum isophotes superimposed on a 
DSS image.

\subsection{Fabry-Perot interferograms and channel maps}

Fig. 2 presents channel maps for H18 with a field-of-view of
is 1.8' x 1.8'. The velocity amplitude is low, with the lowest-velocity
channel at 4003 km s$^{-1}$ and the highest-velocity channel at 4177 km
s$^{-1}$. The optical centers of H18b, c and d are represented by
crosses. Several emission-line regions appear, with velocities between 4003 km s$^{-1}$ and 4068 km s$^{-1}$ south of the principal optical continuum sources.
We can clearly see from channels 4047 to 4090 km s$^{-1}$ 
that H18c and d have a common emission.  North of the group there is
another emission-line region which appears between channels 4068 - 4177
km s$^{-1}$.  Starting at channel 4025 km s$^{-1}$, the emission for H18d
gets continuously stronger, until it reaches channel 4090 km s$^{-1}$.
There are optical counterparts for almost all emission-line regions except the ones between 4003 - 4068 km.s$^{-1}$.
The southern regions between 4003 and 4047 km.s$^{-1}$, form a circular pattern. Looking carefully at optical images (Hickson 1993; Arp 1966) reveals faint
optical counterparts to the southern sources.
This circular structur does not show any evidence for expantion.
Examination of HI channel maps (W\&vG), reveals that the southern emitting regions coincide with HI emission in the range 4015-3994 km.s$^{-1}$.


The extended H$\alpha$
emission of H18 is not concentrated in a single cloud but instead shows substructures. As described in detail below, we do not detect rotation for
individual group members.  Except for the southern and northern clumps,
all the other emission-line regions are aligned along a position angle
of $\sim$ 160$^o$.

\subsection{Velocity field and line-of-sight velocity curves}

Figure 3 presents the velocity field for the H18b,
c and d system. H$\alpha$ intensity isophotes are superimposed on the velocity field.  The total extent of the velocity mapped structure is
$\sim$ 2.2' by 30''.  The radial velocity across the map varies from
4007 km s$^{-1}$, in the southern region, to 4127 km s$^{-1}$, in the
northern and northeastern part of the group. This map shows that there
is no circular motion for the group as a whole. Nevertheless, the
southern region seems to have an independent kinematics from the rest of the
group. \\

 Line-of-sight (LOS) velocity curves (uncorrected for inclination)
can be derived from the velocity field.  Fig. 4 shows LOS velocity
curves along different position angles. Fig. 4a presents
the LOS velocity diagram for the southern region, with a cut along PA
$\sim$ 80$^o$ (see Fig. 1 for the exact location). Fig. 4b presents the LOS velocity curve for the group as a whole, along a PA $\sim$
160$^o$, the PA that best describes the velocity field of the group.

The first plot (Fig. 4a) shows disk-like rotation with a velocity
amplitude of 70 km s$^{-1}$ and an almost solid body motion
across a region of 20". This plot includes all points within a
cone of half radius 20 degrees.  The
emission is rather weak in comparison with the rest of the system. The
measured profiles have a S/N ratio between five and ten. In this region
there are no catalogued members noted by Hickson (1982). This clear
velocity gradient may indicate an independent motion for the southern part
of the system.

Fig. 4b shows the velocity curve along a PA $\sim$ 160$^o$, which is our best estimate for the PA of the major axis of the entire group. For this plot we show the average velocity values with error bars (calculated in a two pixels crown) inside a cone of 20$^o$ half-radius.
 We note that although
there is a large scatter in the velocities (with a dispersion of $\pm$
20 km s$^{-1}$), there is a clear velocity gradient from the northwest
to the southeast of the group, of mean total amplitude of 78 km s$^{-1}$,
over 90 arcsec. We marked on Fig. 4b the positions of the three group
members, H18b, c and d.

\subsection{Monochromatic emission}

Fig. 5
shows isophotes of averaged H$\alpha$ flux (superimposed on the DSS image) 
obtained by integrating the emission above the
continuum. The image was smoothed with a box of 5 $\times$ 5 pixels.
The threshold is 10$^{-16}$ and the step is 5 $\times$ 10$^{-16}$ erg
s$^{-1}$cm$^{-2}$ arcsec$^{-2}$. The total H$\alpha$ flux is 2.5
10$^{-12}$ erg s$^{-1}$cm$^{-2}$.  Using the formula of Osterbrock
(1974) and assuming an electronic density of 1000 cm$^{-3}$, we obtain an
upper limit for the ionized gas mass of M$_{HII}$ = 2 $\times$ 10$^6$
M$_{\odot}$.  The five compact emitting
regions seen in Fig. 5 represent more than 50\% of the total emission
from the source.  As mentioned before, although the gas emission is
clumpy, the profiles are continuous between H18c and H18d. We also note
that regions 2 and 5 do not correspond to any cataloged group members 
(Hickson 1993) although we do detect counterparts for these 
emission regions in our continuum image.

\subsection{Star Formation History of HCG18}

Total H$\alpha$ flux implies a luminosity 
L(H$\alpha$)=2.2 $\times$ 10$^8$ L$_{\odot}$ adopting a distance 
of 54 Mpc for H18 (with H$_o$ = 75 km.s$^{-1}$.Mpc$^{-1}$).

The visual B magnitude in Hickson (1993), yields a 
B luminosity of L(B)=1.1 $\times$ 10$^{10}$ L$_{\odot}$.

From Allam et al. (1996) IRAS fluxes we derive the
total FIR luminosity of L(FIR)=2.42 $\times$ 10$^9$ L$_{\odot}$ for the
whole group.

  The multiwavelenght luminosities (B, H$\alpha$,
FIR) can be used to determine the star formation rates (SFR) for
different epochs in the galaxy history. Table 3 summarizes the different
quantities derived for H18. A recent SFR can be estimated from the B
luminosity and equation 7 in Gallagher et al.  (1984b). This equation
represents the average SFR over the lifetime of the stars which dominate
the blue light (0.4 -- 6 $\times$ 10$^9$ years).  
We then estimate SFR(B)= 0.29 $\times$ 10$^{-10}$ L$_B$ M$_{\odot}$ yr$^{-1}$ (where L$_B$ is the bolometric solar luminosities). And SFR(B)=0.32 $\times$ M$_{\odot}$ yr$^{-1}$.
The current SFR can be deduced from  both FIR (L(FIR))
and H$\alpha$  (L($H\alpha$)) luminosities.  The current SFR is driven by
the number of OB stars. Knowledge of the FIR luminosity may measure this
(through the dust heated by OB stars). But our lack of
understanding of the details of the radiation processes responsible for
the FIR luminosity introduces uncertainties in the calculation of the
recent SFR (Gallagher \& Hunter 1987). The major assumption for the
recent SFR determined using L(FIR) is the role played by the massive
stars in heating the dust.  Thronson \& Telesco (1986) adopted an
SFR(IR)=6.5 $\times$ 10$^{-10}$ L(FIR) M$_{\odot}$ yr$^{-1}$ (with
L(FIR) in L$_{\odot}$) giving the current SFR over 2 $\times$ 10$^6$
years. Under similar assumptions but using a different stellar
luminosity law and different upper mass limits, Gallagher \& Hunter
(1987) deduced another formula which give rates that are half the
values calculated by Thronson \& Telesco (1986) (for a given
luminosity).

The H$\alpha$ luminosity provides another way to estimate the current SFR (Hunter \& Gallagher 1986, Gallagher \& Hunter
1987), by calculating the flux of Lyman continuum photons from hot young stars. In that case, a problem could arise from the effect of dust
extinction and the model chosen to convert the observed L(H$\alpha$) to
Lyman count luminosity (Gusten \& Mezger 1982). Hunter \& Gallagher
(1986) adopted, SFR(H$\alpha$)=7.02 $\times$
10$^{-42}$ L(H$\alpha$) M$_{\odot}$ yr$^{-1}$,  with L(H$\alpha$) in
erg s$^{-1}$.  We found a SFR(FIR)= 1.58 M$_{\odot}$ yr$^{-1}$ and
SFR(H$\alpha$)=6.1 M$_{\odot}$.yr$^{-1}$ for H18. The
factor of four difference between the measurements.  However, is expected given
the assumptions made in the calculations.

%
%

\begin{deluxetable}{llll}
\tablenum{3}
\tablewidth{0pt}
\tablecaption{Derived data}
\tablehead{
\colhead{} & \colhead{HCG 18}}
\startdata
 F(H$\alpha$)erg s$^{-1}$ cm$^{-2}$ & 2.5 10$^{-12}$  \nl
 L(H$\alpha$) L$_{\odot}$            & 2.2 10$^8$   \nl
 L(B) L$_{\odot}$                    & 1.1 10$^{10}$  \nl
 L(FIR) L$_{\odot}$                  &  2.4 10$^{9}$ \nl
 SFR(H$\alpha$) M$_{\odot}$ yr$^{-1}$ &  6.1 \nl
 SFR(B) M$_{\odot}$ yr$^{-1}$         & 0.32 \nl
 SFR(FIR) M$_{\odot}$ yr$^{-1}$       & 1.58  \nl
M$_{HI}$/M$_{HII}$                    & 5000  \nl
M$_{HI}$/L(B) M$_{\odot}$ / L$_{\odot}$ & 0.9  \nl

\enddata

\end{deluxetable}

\section{Comparison with HI data}

   HI data (W\&vG) represents an important piece of information for understanding the nature of H18. They presented HI velocity field and integrated images for
H18 and suggested that it is a single giant cloud (3.0' $\times$ 2.5')
not associated with any individual member of the group.  They found evidence for systematic rotation along a
position angle of 20$^o$ and velocities amplitude from 150 km s$^{-1}$. Inspection of their Fig. 3, which presents the HI velocity
field of the group, shows a variation of the kinematic position angle
along the major axis of the velocity field. 

   In order to compare W\&vG's data with the H$\alpha$ kinematics we
performed a gaussian smoothing of our data cube with FWHM = 22". In
order to check for possible smoothing artifact due to very bright region
(H18d), we normalized the ionized gas data cube and found no
contamination was present. Fig. 6 shows the resultant H$\alpha$ velocity field superimposed on the HI velocity field for comparison.  The velocity amplitude is between 170 km s$^{-1}$, consistent with the HI velocity range.

   The two velocity fields present similarities and differences.
The global appearance of the H$\alpha$ velocity field appears to be
less regular than that for the HI.  The H$\alpha$ map shows a region of low
velocities coinciding with H18d and not present in HI.

   The southern part of the H$\alpha$ and HI velocity fields are
similar.  In this region the HI velocities range from 4020-4050 km s$^{-1}$ while the H$\alpha$ velocities go from 4037 km s$^{-1}$ to 4072 km s$^{-1}$.

   The HI and H$\alpha$ velocity fields show less the
agreement in the northern part of H18.  In the center, at the 4050 km s$^{-1}$, the PA of the major axis of the H$\alpha$ velocity field is
close to 0 $\pm$ 5$^o$  while it is close to 20 $\pm$ 5$^o$ for HI.  In the northern part, the morphology of the two maps is different mainly due to the presence of the low velocity region to the west of the H$\alpha$ velocity field. In the northeast, where the influence of the low velocity region is not
visible, the coincidence with the HI isovelocities is better.

 Fig. 4c presents the LOS velocity diagram constructed from the
smoothed velocity field at PA=20$^o$.  The center was chosen to coincide
with H18c. We also include the data points from W\&vG derived from their Fig 3.  Our velocity amplitude $\sim$ 150 km s$^{-1}$ but the rotation while systematic is obviously not axisymmetric.
The HI velocities, in comparison suggest a solid-body
rotation. The velocity amplitude is close to the value found by us and
the gradient is slightly steeper.

\section{Discussion -- The nature of H18}

  The nature of H18 in light of HI observations was discussed by W\&vG.
They considered two possibilities for the nature of this system: 1)
it is a knotty irregular galaxy (hereafter referred to as ``the 
irregular-galaxy scenario''), or 2)
it is an interacting group and the observed HI cloud is a remnant of
an interpenetrating collision which stripped the gas from the colliding
galaxies. Our data and analysis confirm the first scenario.

\subsection{Arguments in favour of the irregular-galaxy scenario}

 The strongest argument in favour of H18 being an irregular galaxy
comes from the kinematics of the HI and H$\alpha$ gas.  The velocity
maps show velocity amplitudes of $\pm$ 70 km s$^{-1}$ and gradients of
15km s$^{-1}$ kpc$^{-1}$ for the ionized gas and 10km s$^{-1}$ kpc$^{-1}$
for the HI.  These confirm that H18 is a slow rotator.  The few detailed
studies concerning the kinematics concerning irregular galaxies (Hunter 1982, Tomita et al. 1998) confirm that irregular galaxies are indeed slow rotators.

   Hunter (1982) studied a sample of 15 irregular galaxies, five of
which showed velocity gradients between 60 and 80 km s$^{-1}$ kpc$^{-1}$, consistent
with results found for H18.  Tomita et al. (1998) produced
position--velocity diagrams for four dwarf galaxies along several slit
orientations.  They did not detect disk-like rotation but velocity gradients
were clearly seen.  They also found a velocity difference of 10  to 20 km s$^{-1}$ between the HII regions and the HI gas disk.  Saito et al. (1992) reported a kinematic study of the
ionized gas in IC 10 and, particularly, a comparison between the ionized
and HI gas kinematics for this galaxy giving the same kind of velocity differences that we found for H18.  Two other works have studied a
few irregular galaxies with a Fabry Perot interferometer.
Sasaki et al. (1997) reported observations of NGC 4449 and its
H$\alpha$ velocity field that show a kpc-scale mosaic structure
of blueshift and redshift components with a slow global rotation. They confirmed the counter rotation between ionized gas and
the HI halo for this galaxy. Rosado et al. (1998) and Valdez \& Rosado
(1998) present an optical velocity field for NGC 4449 showing a
decreasing gradient along the optical bar and an anticorrelation with
respect to the HI velocity field.

The LOS velocity curve of the southern part (Fig. 4a) suggests an independent disk rotation. The velocity curve along the major axis of H18
(Fig 4b) shows that H18b and c lie on the curve, suggesting that
they are gravitationaly bound.  

 The ratio L$_B$/L(H$\alpha$) of H18 is
consistent with the irregular-galaxy hypothesis. 
Hunter et al. (1989) find a value for the ratio L$_B$
/L(H$\alpha$) $\sim$ 44, for giant irregular galaxies while we find
50 for H18.  

  One last piece of evidence for the irregular-galaxy scenario comes from
the HI data.  The size of the large HI cloud around H18 measured by
W\&vG and the total HI mass they found (M(HI)=10$^{10}$ M$_{\odot}$)
are consistent with values found for HI clouds around irregular galaxies
such as IC 10 (Shostak \& Skillman 1989) and NGC 4449 (Hunter et al.
1998). Normalized by the total B luminosity the HI total mass also
compares with the typical values found for the Hunter (1993) sample.

\subsection{Differences between H18 and irregular galaxies}

 The main difference between H18 and irregular galaxies concerns the FIR
properties and star formation rates.  Hunter et al. (1989, 1993)
present properties (in FIR, H$\alpha$ and broad band imaging) for a sample
of 43 irregular galaxies of different types (dwarf, giants, distant,
amorphous). We used their sample of dwarf and giant irregular galaxies  
as a control sample for comparison with the properties of H18.

  We measured ratios of
L(FIR)/L$_B$ = 0.22 and L(FIR)/L(H$\alpha$)=11 for H18.  The L(FIR)/L$_B$
ratio is significantly lower than the mean value of 1.9  and 1.0 found for
Hunter et al.'s (1989) sample of giant and dwarf irregulars respectively
(the  two other classes, ``amorphous'' and ``distant'', show much larger
ratios).  The L(FIR)/L(H$\alpha$) ratio is also lower than the mean
values of 90 and 71 for the giant and dwarf irregulars respectively.
However, within each subclass, giant, dwarf, amorphous and distant,
there is significant scatter in the ratios.  We also derived a dust
temperature for H18 (from the S(100$\mu$) and S(60$\mu$) IRAS data)
of T$_d$=27K, which is cooler than typical dust temperatures derived
for other irregular galaxies.

   The recent SFR (SFR(B)) and the current SFR ( SFR(FIR) or
SFR(H$\alpha$) ) are much higher (by one or two orders of magnitude)
compared with values from Hunter et al's sample.  If, however, we
calculate a SFR per area, using the SFR(H$\alpha$) and taking the area
to be that enclosed within the ellipse of major and minor axes defined
by the H$\alpha$ integrated flux map (Fig.5), we find that the SFR/area is 5 $\times$
10$^{-9}$ M$_{\odot}$ yr$^{-1}$ pc$^{-2}$. Hunter \& Gallagher (1986)
have a similar average for the SFR/area for the giant irregular galaxies
of their sample.  The other subclasses of irregular galaxies have average values
for the SFR/area ratio that are slightly lower than that found for H18.

\section{Summary and final remarks}

HCG 18 is a special case in the Hickson compact group catalogue
(Hickson, 1982).  It was originally thought to be a group composed of
three members, H18b, c and d and a discordant-redshift galaxy, H18a.
The three members are not well separated and they are embedded in a
diffuse halo. A key question raised by W\&vG was about the nature of
H18.  Our H$\alpha$ observations suggest that H18 is in fact a giant
irregular galaxy confirming previous HI results.

    Evidence in favour of the irregular-galaxy scenario here. We find that the blobs (previously called galaxies, when
the system was classified as a group) are kinematically connected and
therefore they may form a single object.  We find a velocity gradient
similar to what was found by W\&vG from the HI data and consistent
with values for irregular galaxies. A comparison between the H$\alpha$ and the
HI velocity fields shows a difference on their morphologies but the
velocity amplitudes and gradients are similar. 
Various authors also show differences
between the HI and the optical kinematics for other irregular galaxies
(see Section 5.1).  The H$\alpha$ emission shows that the ionized gas
distribution is clumpy which is rather common for irregular galaxies. 
The values for the total luminosities in B, H$\alpha$ and FIR are larger
than those for the Hunter et al.'s sample (1989) of irregular galaxies
but if we normalize the L(H$\alpha$) by the total B luminosity or the
L(FIR) by the total B luminosity, it appears that the ratios are close
to what other authors find for giant irregular galaxies.
 The inferred
star formation rate of H18 is higher than the average for irregular
galaxies although the values are comparable when they are normalized by the area
of the galaxies.

  Two others groups that may be classified as a
single object, an irregular galaxy, are HCG 31 and HCG 54.  H54 is formed by
three members classified as irregulars and one late spiral and H31 is formed of four members, a and c seems to be an single object.  Kinematical
study is necessary to confirm if it is an irregular galaxy like H18.
If true, this would suggest that only a small fraction ($\sim$ 3\%)
of catalogues of compact groups may be misclassified single objects.

\acknowledgments

The authors would like to thank the CFHT staff for helping during
the observations and Drs. R. Sancisi, M. Moles, D.A Hunter and the anonymous referee for helpful
discussions and corrections. H. Plana acknowledges the financial support of the Brazilian
FAPESP, under contract 96/06722-0 and the Mexican CONACYT.

%
%

\begin{figure*}
\figurenum{1}
\plotfiddle{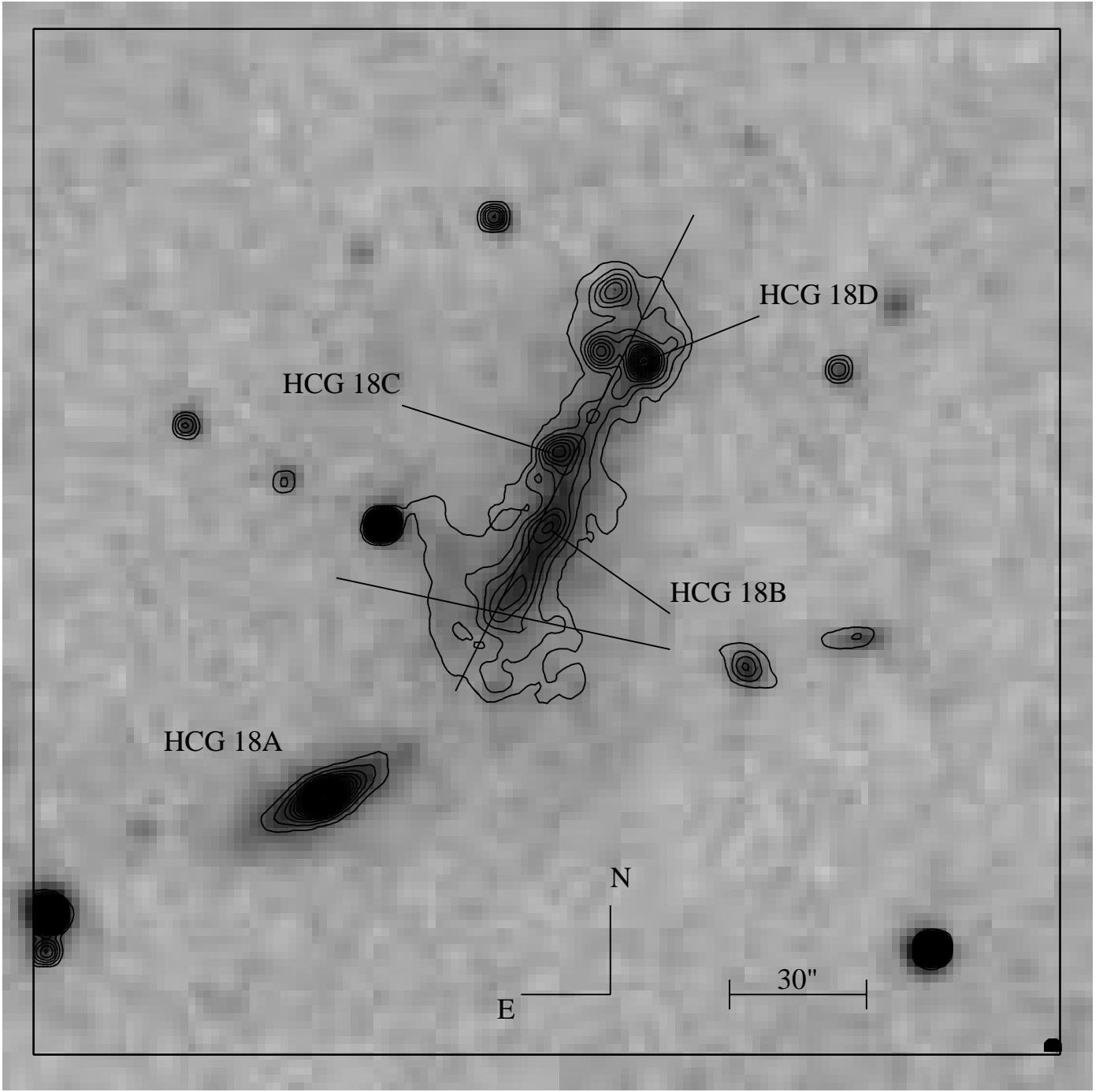}{16cm}{0}{100}{100}{-300}{-150}
\caption{Continuum image of H18a, b, c and d. 
The contours (in arbitrary units) were plotted
after a rectangular smoothing with a box of 5 $\times$ 5 
pixels. The continuum map has been superimposed to a DSS image. We also show the
three directions, one east - west (on the southern region), one north - south
along the group major axis and one coinciding with the HI major axis
(PA=20$^o$). These axis were used for deriving the LOS velocity curves shown
in Fig.4.}
\end{figure*}

\clearpage

%
%

\begin{figure*}
\figurenum{2}
\plotone{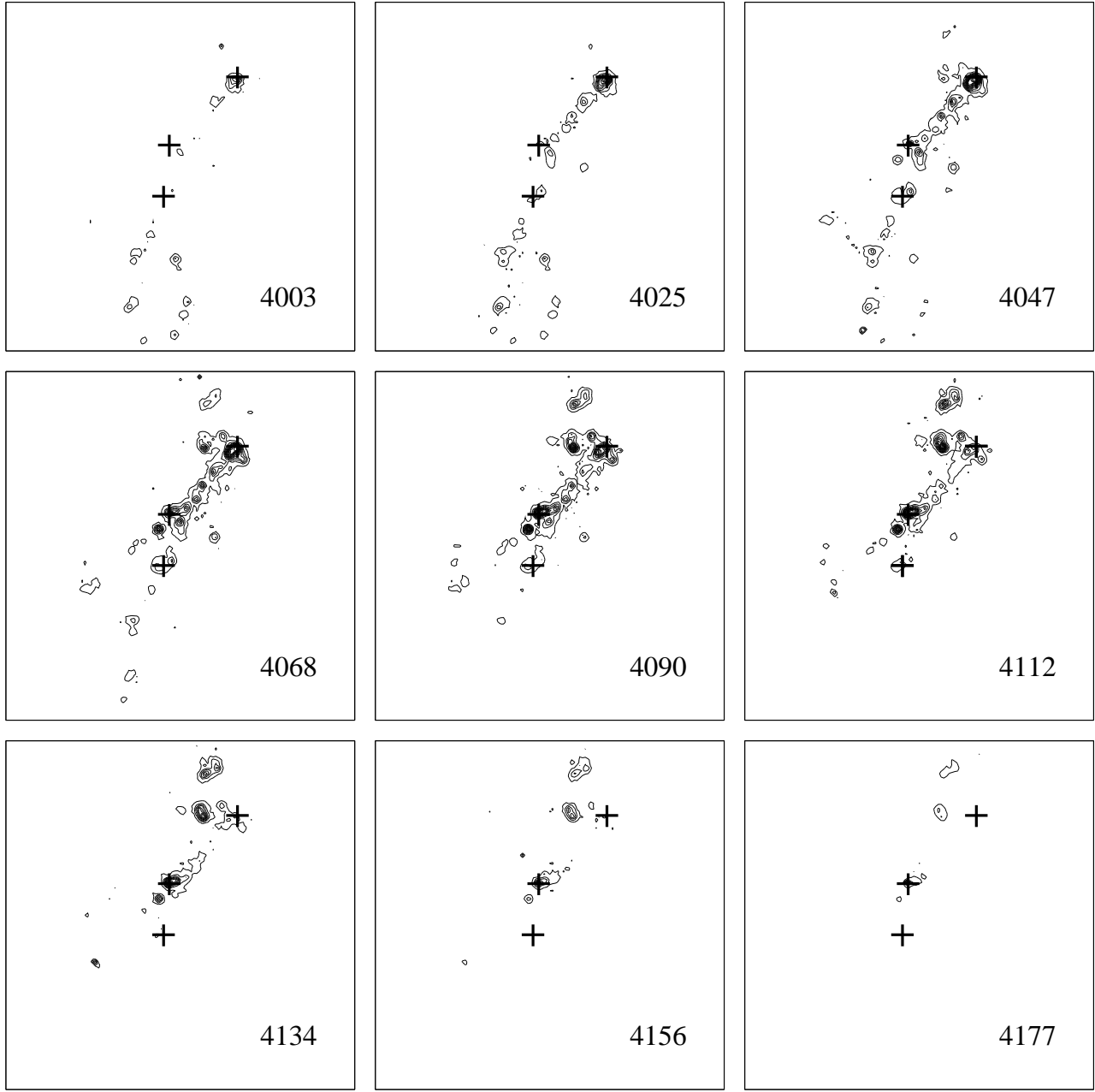}
\caption{Channel maps of H18. We present 9 of the 24 channels we have scanned.
The field of view is 1.8' x 1.8', North is up and East is to the left.
Crosses represent the positions of H18b, c and d . In each pannel we mark the
velocity of the channel. The lowest isointensity represents a flux density
of 4.4 10$^{-16}$ erg s$^{-1}$cm$^{-2}$ arcsec$^{-2}$. }
\end{figure*}

\clearpage
%
%

\begin{figure*}
\figurenum{3}
\plotfiddle{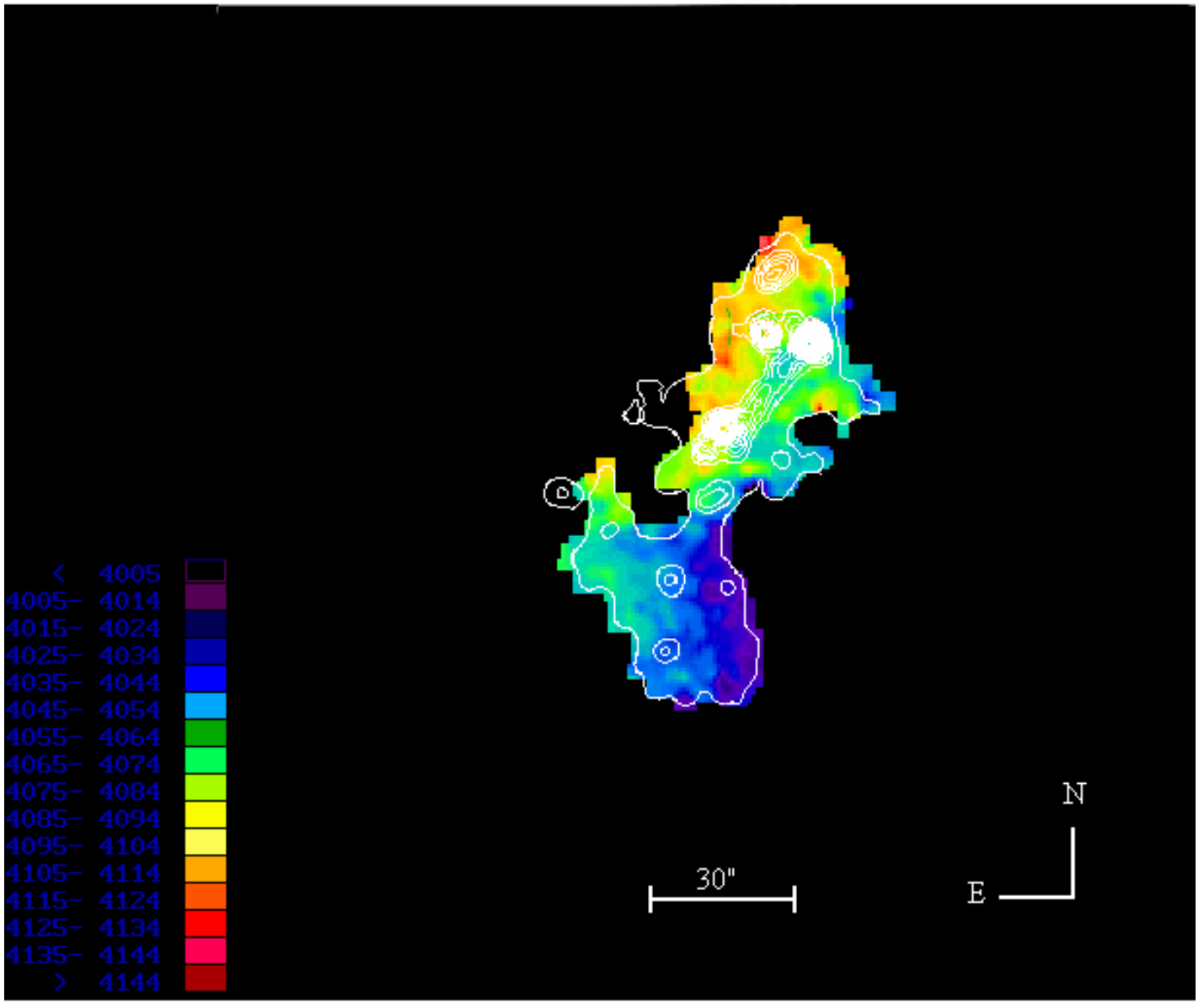}{16cm}{0}{87}{87}{-260}{-50}
\caption{Velocity field of H18. We present the velocity map of the group and,
superimposed with white isophote,  the H$\alpha$ integrated flux map  (See Fig. 5). Velocity map has been spacialy smoothed with a box of 3$\times$ 3
pixels. }
\end{figure*}

\clearpage

%
%

\begin{figure*}
\figurenum{4}
\plotfiddle{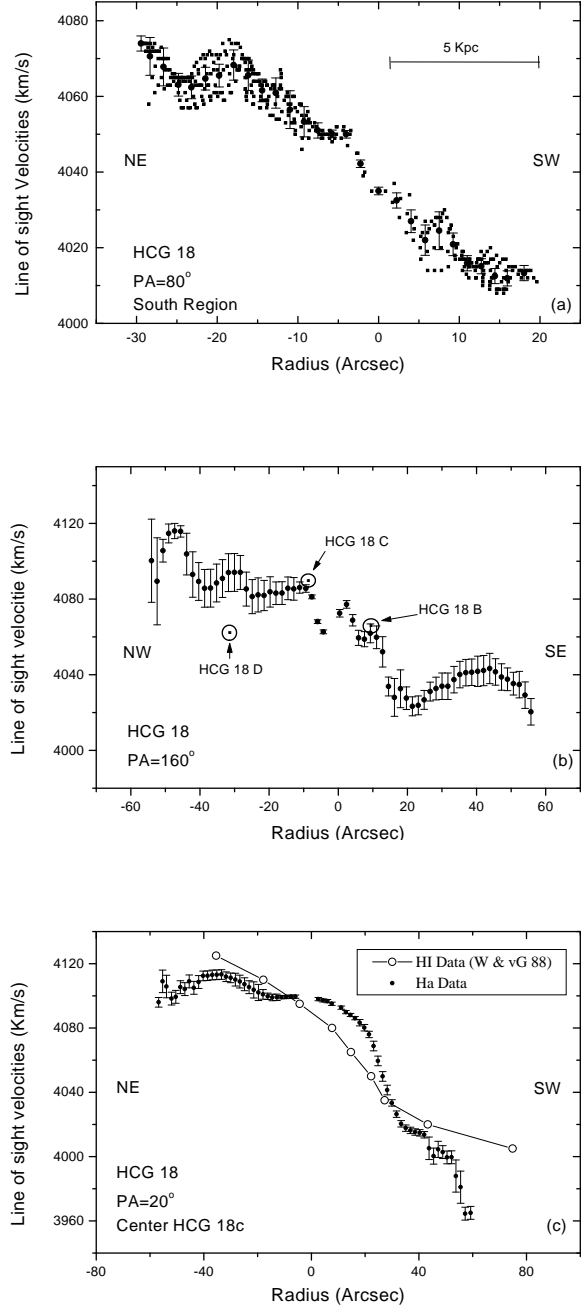}{17.5cm}{0}{85}{85}{-250}{-50}
\caption{Line of sight velocity curves of HCG 18. In all plots, filled
 circles represent the average values of velocities
in a 2*pixels crown with the dispersion error bars indicated.
(a) Velocity curve of the southern region of the group, along an axis with
 a PA=80$^o$ (see fig. 1 for the exact location of the axis). The plot shows
 all the points inside a cone with an half angle at the summit of 20$^o$.
(b) Line of sight velocities along the major axis of
HCG 18 (PA=160$^o$). The positions of the three group members, HCG 18b, c and d are marked .
(c) This plot shows the velocity curve across an axis with a
 PA=20$^o$ using the velocity field presented on fig. 6 (at the resolution
 of HI data). Overplotted are the data points taken from the VF of
 Williams \& van Gorkom (1988) along the same PA of 20$^o$.}
\end{figure*}

\clearpage

%
%

\begin{figure*}
\figurenum{5}
\plotfiddle{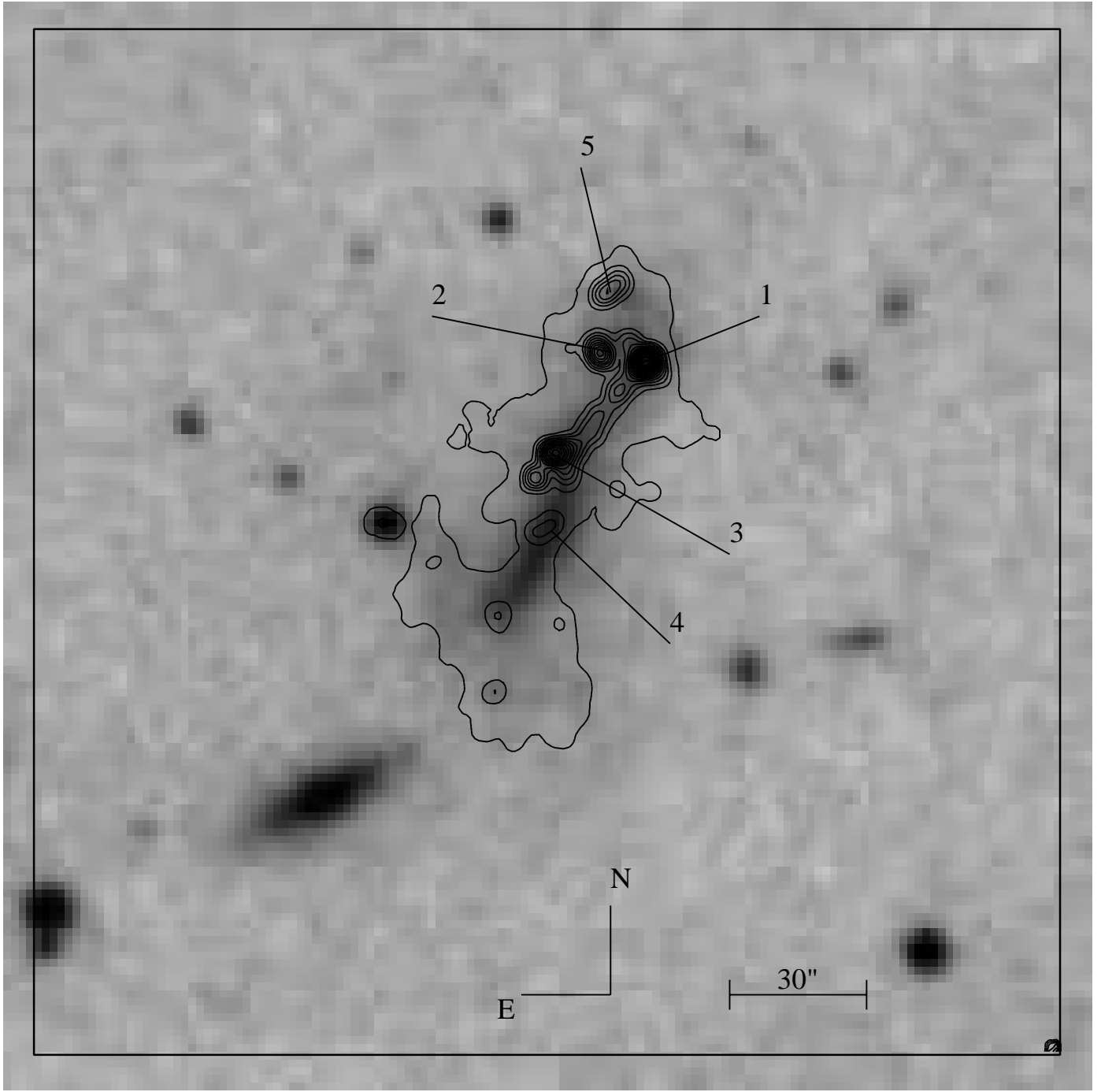}{16cm}{0}{100}{100}{-300}{-150}
\caption{ the H$\alpha$ integrated flux map of HCG 18. The map has been spatially smoother
with a rectangular box of 5 x 5 pixels. Isointensities are in unit of 10$^{-16}$ erg s$^{-1}$cm$^{-2}$ arcsec$^{-2}$. The lowest level is 1.1 and step the is 5. The monochromatic image has been superimposed to a DSS image. }
\end{figure*}

\clearpage

%
%

\begin{figure*}
\figurenum{6}
\plotfiddle{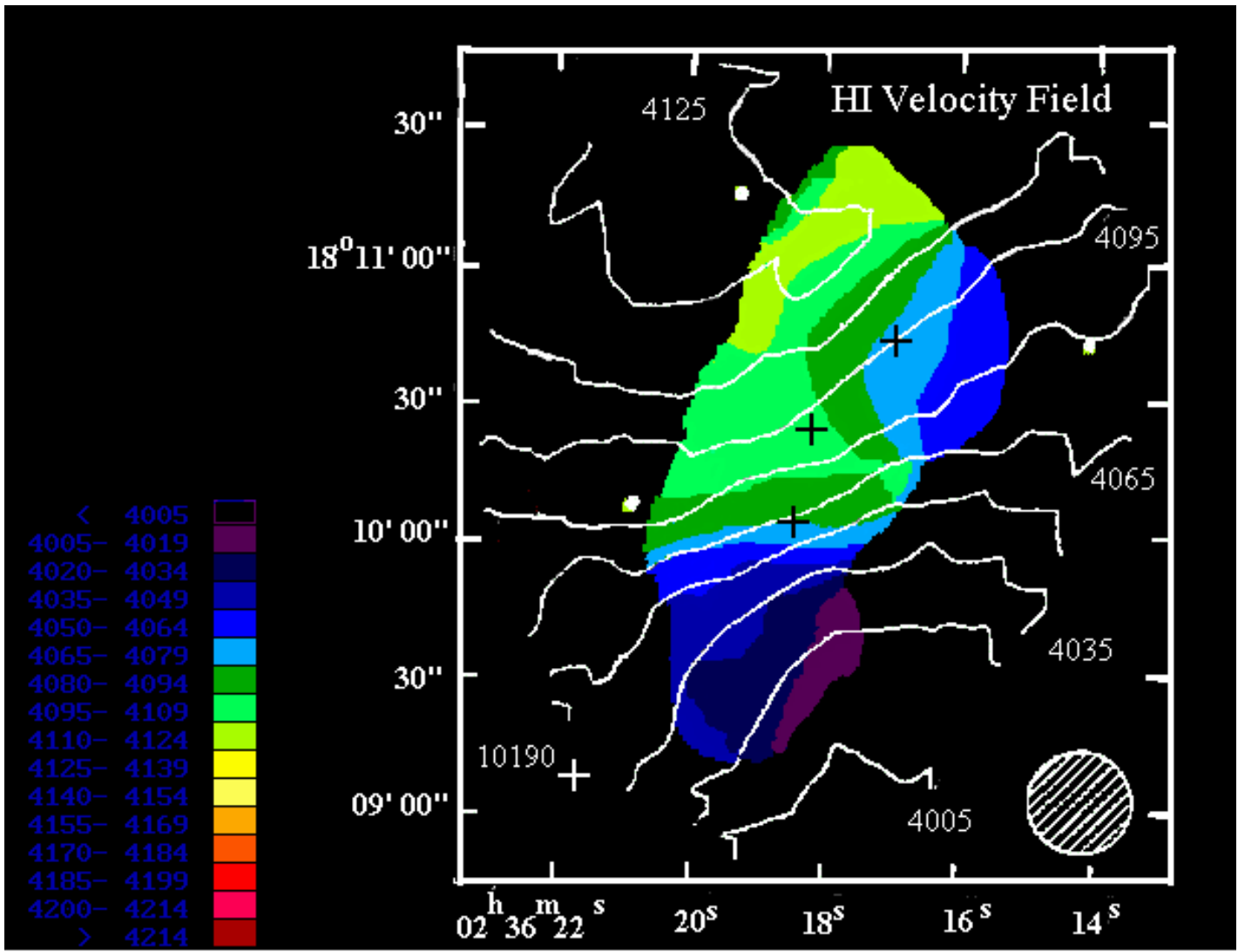}{14cm}{0}{87}{87}{-300}{-200}
\caption{Superposition of the HI velocity map from Williams \& van Gorkom 1988 on our H$\alpha$ VF. In order to match the resolution of the two maps, we
performed a gaussian smoothing of 22" (see text). The crosses represent
positions of group members b, c and d.}
\end{figure*}

\end{document}